\documentstyle[12pt]{article}
\begin{document}
\centerline {\Large {\bf High Reproduction Rate versus Sexual Fidelity}}

\bigskip
\centerline {\bf A.O. Sousa and S. Moss de Oliveira} 

\bigskip
\centerline {\it Instituto de F\'{\i}sica, Universidade Federal 
Fluminense} 

\centerline{\it Av. Litor\^anea s/n, Boa Viagem, Niter\'oi 24210-340, RJ, 
Brazil}

\bigskip
\noindent {\bf Abstract}

\bigskip
We introduce fidelity into the bit-string Penna model for biological 
ageing and study the advantage of this fidelity when it produces  
a higher survival probability of the offspring due to paternal care. 
We attribute a {\it lower reproduction rate} to the faithful males but a {\it 
higher 
death probability} to the offspring of non-faithful males that abandon  
the pups to mate other females. The fidelity is considered as a 
genetic trait which is transmitted to the male offspring (with or without error).  
We show that {\it nature may prefer a lower reproduction rate} to warrant 
the survival of the offspring already born.  

\section{Introduction}

It is not difficult to find in the animal kingdom species that live and work in 
sexual pairs, but sometimes 
have extra-pair relations. Biologists believe that these pairs are formed 
in order to better take care of the pups, and that the extra-pair 
relations have the genetic purpose to maximize the variability 
of their offspring or to produce some fitness benefit for them [1]. 
The Scandinavian great reed warbler is one of the 
species that presents these extra-pair matings.   
However, independent of its origin (social or genetic), true monogamy seems to be 
rare in Nature.

The Penna model for biological ageing [2] is a Monte Carlo simulation technique  
based on 
the mutation accumulation hypothesis. It has successfully reproduced many 
different characteristics of living species, as the catastrophic senescence 
of pacific salmon [3], the inheritance of longevity [4] and the self-organization 
of female menopause [5]. The extra-pair relations mentioned above 
have also been studied through this model [6]. Martins and Penna have obtained 
that the 
offspring generated by extra-pair relations are genetically stronger and 
present a higher survival probability than those generated by social relations. 

In this paper we are interested in using the Penna model to study the true  
monogamy, rarely found in 
Nature. One example is the California mouse. In this species a female is 
not able to sustain one to three pups alone. The pups are born at 
the coldest time of the year and depend on the parents body heat to survive. 
According to the biologist David Gubernick, as cited in Science [1], the situation 
is so dramatic 
that if the male leaves or is taken away, the female abandons or kills the pups. 
However, he also points out that other species of mice that live in the same 
environment are promiscuous. That is, the reason for true monogamy is 
still an open question under study. We have adopted the strategy of considering 
monogamy as a genetic trait, exclusively related to paternal care. Our assumption 
that male fidelity is genetically transmitted is analogous to the recent findings 
that the gene {\it Mest} regulates maternal care [7,8].  

In the next section we explain the Penna model and how fidelity is introduced.  
In section 3 we present our results and in section 4 the conclusions.

\section{The Sexual Penna model and Fidelity}

We will now describe the sexual version of the Penna model;  
details and applications can be found, for instance, in references [5,9]. 
The genome of each individual is represented by two bit-strings of 32 bits that are 
{\it read in parallel}; that is, there are 32 positions to be read, to each 
position 
corresponding two bits. One time-step corresponds to read one position of all the  
genomes. In this way, each individual can live at most for 32 time-steps 
(``years''). Genetic diseases are represented by bits 1. If an individual 
has two bits 1 (homozygotous) at the third position, for instance, it starts to 
suffer from a genetic disease at its third year of life. If it is an homozygotous 
position 
with two bits zero, no disease appears at that age. If the individual is 
heterozygotous in 
some position, it will get sick only if that position is a dominant one. The 
number of dominant genes and its randomly chosen positions are defined at   
the beginning of the simulation; they are 
the same for all individuals and remain constant. When the number of accumulated 
diseases of any individual reaches a threshold $T$, the individual dies.  

The individuals may also be killed due to a  
lack of space and food, according to the logistic Verhulst factor 
$V=1-N(t)/N_{max}$, where $N(t)$ is the current population size and $N_{max}$ the 
carrying capacity of the environment. At every time step and for each individual 
a random number between zero and 1 is generated and compared with $V$: if this  
number is greater than $V$, the individual dies independently of its age or 
number of accumulated diseases. 

If a female succeeds in surviving until the minimum reproduction age $R$, 
it generates, with probability $p$, $b$ offspring every year. The female randomly 
chooses a male to mate, the age of which must also be greater or equal to $R$. 
The offspring genome is constructed from the parents' ones; firstly the strings 
of the mother are randomly crossed, and a female gamete is produced. $M_m$ 
deleterious mutations are then randomly introduced. The same process occurs with 
the father's genome (with $M_f$ mutations), and the union of the two remaining 
gametes form the new genome.  
Deleterious mutation means that if the randomly chosen bit of the parent genome 
is equal to 1, it remains 1 in the offspring genome, but if it is equal to zero 
in the parent genome, it is set to 1 in the baby genome. It is well known 
[5,9,10]  that due to the dynamics of the model, the bits 1 accumulate, 
after many generations, at the end part of the genomes, that is, after the 
minimum reproduction age $R$. For this reason ageing appears: the survival 
probabilities decrease with age. The sex of the baby is randomly chosen,
each one with probability $50\%$.

Let's see now how fidelity is introduced. We assume that if a female reproduces 
this year, she spents the next two following years without reproducing. So we 
consider two time steps as the {\it parental care period}. Remembering that in our 
simulations the female choses the male, if the male is 
a faithful one, he will refuse, during this period, to mate any female that 
eventually choses him as a partner. The non-faithful male accepts any invitation, 
but his offspring still under parental care pay the price for the abandonment: they 
have an extra probability $P_d$ of dying. The male offspring of a faithful father 
will also be faithful, with probability $P_f$. This means that if the father is 
faithful and $P_f = 1$, the male offspring will necessarily be faithful. 
$P_f$ is also the probability of a non-faithful male having a non-faithful
offspring.

\section{Results} 

{\it We start our simulations with half of the males faithful and half 
non-faithful}.  
In Fig.1 we show the final percentages (after many generations) of faithful
males as a function of the offspring death probability $P_d$, for the 
cases where the male offspring inherits the father's fidelity state with 
probability $P_f =~1$ (full line) and with probability $P_f = 0.8$ 
(dashed line). This last case means that the offspring of a faithful 
father has a $20\%$ probability of being non-faithful and vice-versa. 
From this figure we can see that as the death probability of 
the abandoned pups increases, the percentage of faithful fathers 
increases. From the solid curve it is easy to notice that there is a compromise 
between the lower reproduction rate of the faithful males and the death  
probability of 
the already born offspring abandoned by father: if $P_d < 0.3$, a high 
reproduction rate dominates and after many generations the    
faithful males disappear from the population. However, for 
$P_d = 1$ the opposite occurs, since there is a strong selection 
pressure against the non-faithful males to warrant the survival of 
the already born offspring. 

From the dashed curve ($P_f = 0.8$) it can be seen that for $P_d = 0$ a high 
percentage (greater than $20\%$) of faithful males remains in the final 
population. The reason is that for $P_d=0$ there is no selection pressure.
There is a probability that non-faithful males, which have a high 
reproduction rate, generate faithful offspring; these offspring 
are introduced into the population and, without any pressure, remain there.
At this point ($P_f=0.8$ and $P_d=0$) we have computed which percentages 
of faithful males descend from faithful and non-faithul fathers. We have 
obtained that for the $26.95\%$ of faithful males that remain in the population, 
$9.28\%$ of them descend from faithful fathers and $17.67\%$ from non-faithful 
ones.    

In Fig.2a we present the time evolution of the populations for $P_f = 1$, 
and in Fig.2b for $P_f = 0.8$. The inset show the final population sizes 
as a function of $P_d$. From Fig.2a it can be seen that the population sizes 
decrease until $P_d=0.4$ and then increase for increasing values of $P_d$, 
stabilizing around the same population size of $P_d=0.3$.    
For $P_f = 0.8$ (Fig.2b) the population sizes decrease until $P_d=0.7$, and then 
stabilize around the same final size for increasing values of $P_d$.

Fig.3 shows the survival rates for $P_f = 1$ and $P_d = 0.0$ (circles), 0.5 
(squares) and 0.9 (triangles). It can be noticed that for $P_d = 0.5$ the 
child mortality is greater, since $P_d$ is already large and nearly $50\%$ of 
the males (see fig.1, solid curve) are not faithful. The results obtained for 
$P_f=0.8$ are similar. 

The survival rate is defined, for a stable population, as the ratio 
$$S(a)=N(a+1)/N(a) \,\,\, ,$$
where $N(a)$ is the number of individuals of age 
$a$. A stable population means that the number of individuals of any given 
age $a$ is constant in time. It is important to emphasize that all 
curves presented here correspond to already stable situations. To 
obtain each of them  we simulated 20 different populations (samples) 
during 800,000 time steps, and averaged the final results. The 
parameters of the simulations are:

\bigskip
\noindent Initial population = 20,000 individuals (half for each sex);

\noindent Maximum population size $N_{max} = 200,000$;

\noindent Limit number of allowed diseases $T = 3$;

\noindent Minimum reproduction age $R = 10$; 

\noindent Probability to give birth $p = 0.5$; 

\noindent Number of offspring $b = 2$; 

\noindent Number of mutations at birth  $M_m = M_f = 1$;

\noindent Number of dominant positions = 6 (in 32). 

\section{Conclusions}

We have used the Penna bit-string model for biological ageing to study 
the problem of true monogamy, rarely found in Nature. In our simulations 
a female that gives birth necessarily waits two time steps before giving 
birth again. We call this period the parental care period. A faithful father  
also cannot reproduce during this period, but a non-faithful one can accept any 
female that randomly choses him to mate, abandoning the pups already born.
The abandoned pups have, as a consequence, an extra probability to die. 
In this way there is a competition between the reproduction rate and 
the death probability of already born pups. We show that depending on 
this death probability, nature may prefers a lower reproduction rate to 
warrant the survival of those babies already born. We consider the paternal 
fidelity an expression of paternal care, and so admit it as a genetic trait 
to be transmitted to the male offspring.  

\bigskip
\noindent Acknowledgements: to P.M.C. de Oliveira and D. Stauffer for 
important discussions and a critical reading of the manuscript; to CNPq, CAPES 
and FAPERJ for financial support.     
  
\newpage

\noindent {\Large \bf References}

\begin{description}

\item [ 1-] V. Morell, {\it Science} {\bf 281}, 1983 (1998).
\item [ 2-] T.J.P. Penna, {\it J.Stat.Phys.} {\bf 78}, 1629 (1995).
\item [ 3-] K.W. Wachter and C.E. Finch, {\it Betwee Zeus and the Salmon. The 
Biodemography of Longevity}, National Academy Press, Washington D.C.;  
T.J.P. Penna, S. Moss de Oliveira and D. Stauffer, {\it Phys.Rev.} 
{\bf E52}, 3309 (1995).  
\item [ 4-] P.M.C. de Oliveira, S. Moss de Oliveira, A.T. Bernardes 
and D. Stauffer, {\it Lancet} {\bf 352}, 911 (1998).
\item [ 5-] S. Moss de Oliveira, P.M.C. de Oliveira and D. Stauffer, 
{\it Evolution, Money, War and Computers}, Teubner, Sttutgart-Leipzig (1999). 
\item [ 6-] S.G.F. Martins and T.J.P. Penna, {\it Int.J.Mod.Phys.} {\bf C9}, 491 
(1998).
\item [ 7-] L. Lefebvre, S. Viville, S. C. Barton, F.  
Ishino, E.B. Keverne and M.A. Surani, {\it Nature Genetics} {\bf 20}, 163 
(1998).
\item [ 8-] R.S. Bridges, {\it Nature Genetics} {\bf 20}, 108 (1998). 
\item [ 9-] A.T. Bernardes, {\it Annual Reviews of Computational Physics IV}, 
edited by D. Stauffer, World Scientific, Singapore (1996).  
\item [10-] J.S. Sa Martins and S. Moss de Oliveira, {\it Int.J.Mod.Phys.} 
{\bf C9}, 421 (1998).   

\end{description}

\newpage
\centerline{\bf Figure Captions}

\bigskip
\noindent Fig.1 - Final percentages of faithful males in the population as a 
function of the death probability of the abandoned pups. The solid line  
corresponds to the cases where the offspring fidelity state is the same of 
the father. The dashed line corresponds to the cases where the offspring 
inherit the same fidelity of the father with probability $80\%$. 

\bigskip
\noindent Fig.2a - Time evolution of the populations (linear-log scale) for 
$P_f=1$ and different offspring death probabilities $P_d$. The inset shows 
the final population sizes as a function of $P_d$. For $0.6 \le P_d \le 1$ 
the final sizes are all very close to that for $P_d = 0.3$.

\bigskip
\noindent Fig.2b - The same as Fig.2a for $P_f = 0.8$. 

\bigskip
\noindent Fig.3 - Survival rates as a function of age for $P_f = 1$ and 
different values of $P_d$; circles correspond to $P_d = 0.1$, squares 
to 0.5 and triangles to 0.9. A higher child mortality can be noticed 
for $P_d = 0.5$.    
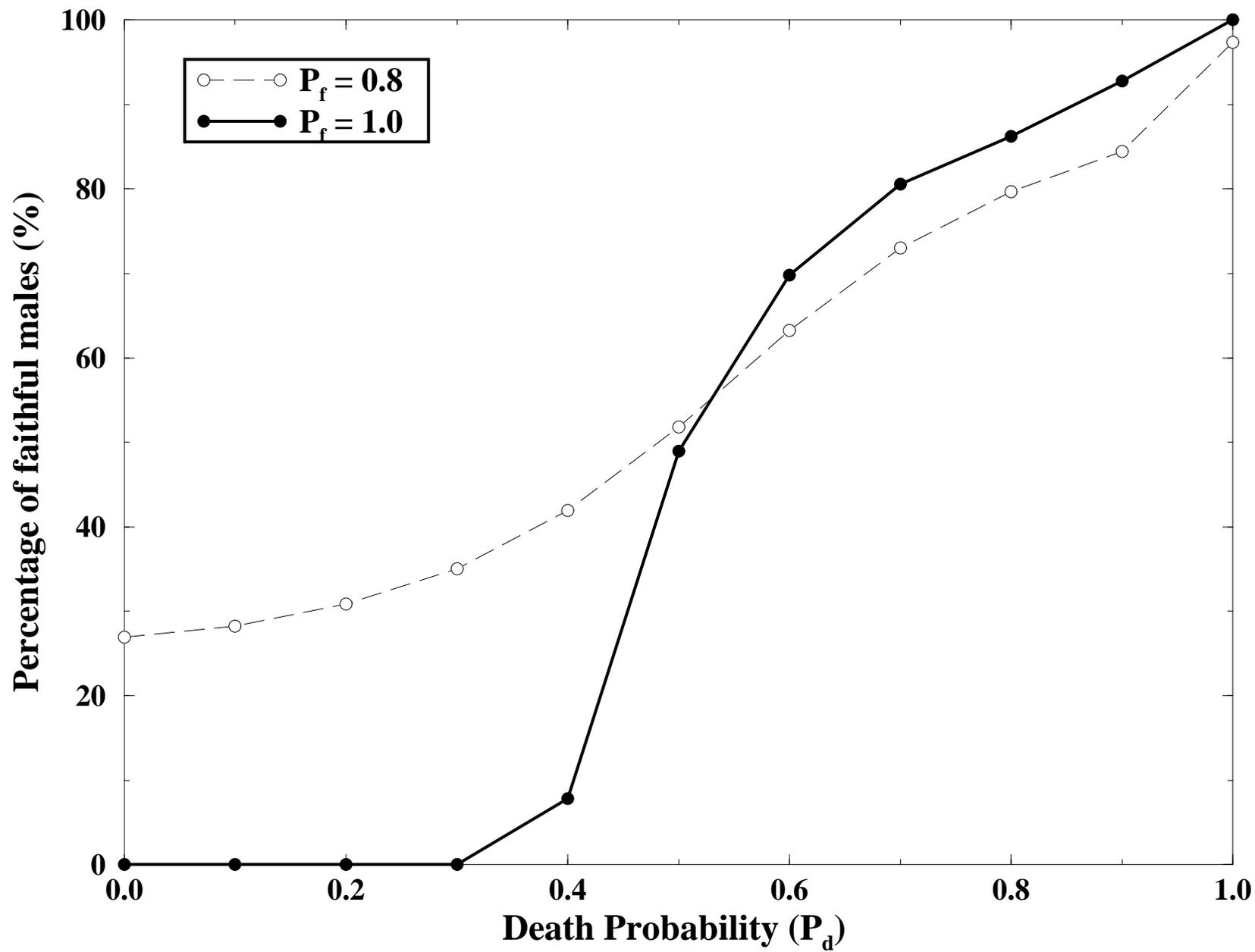
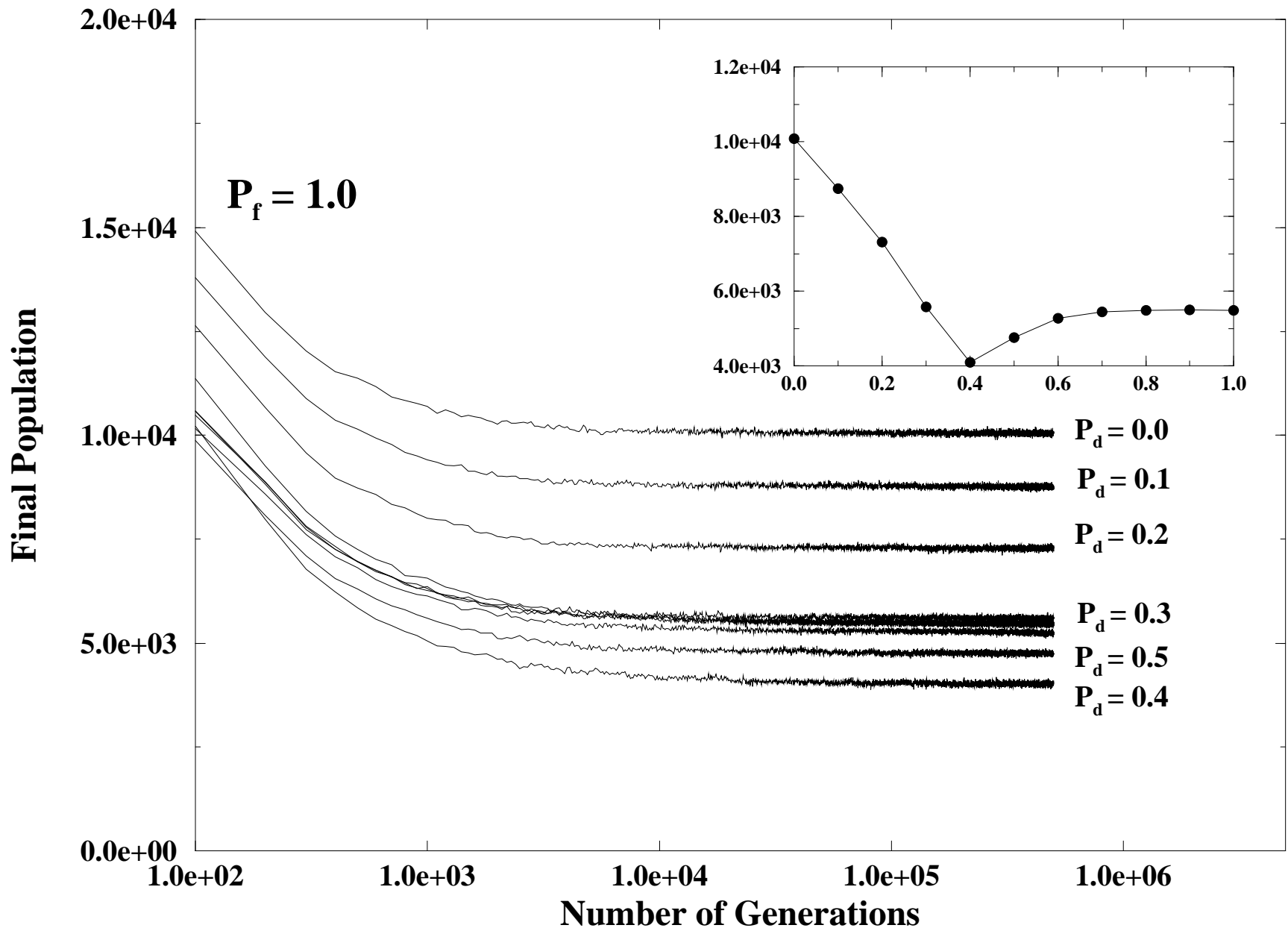
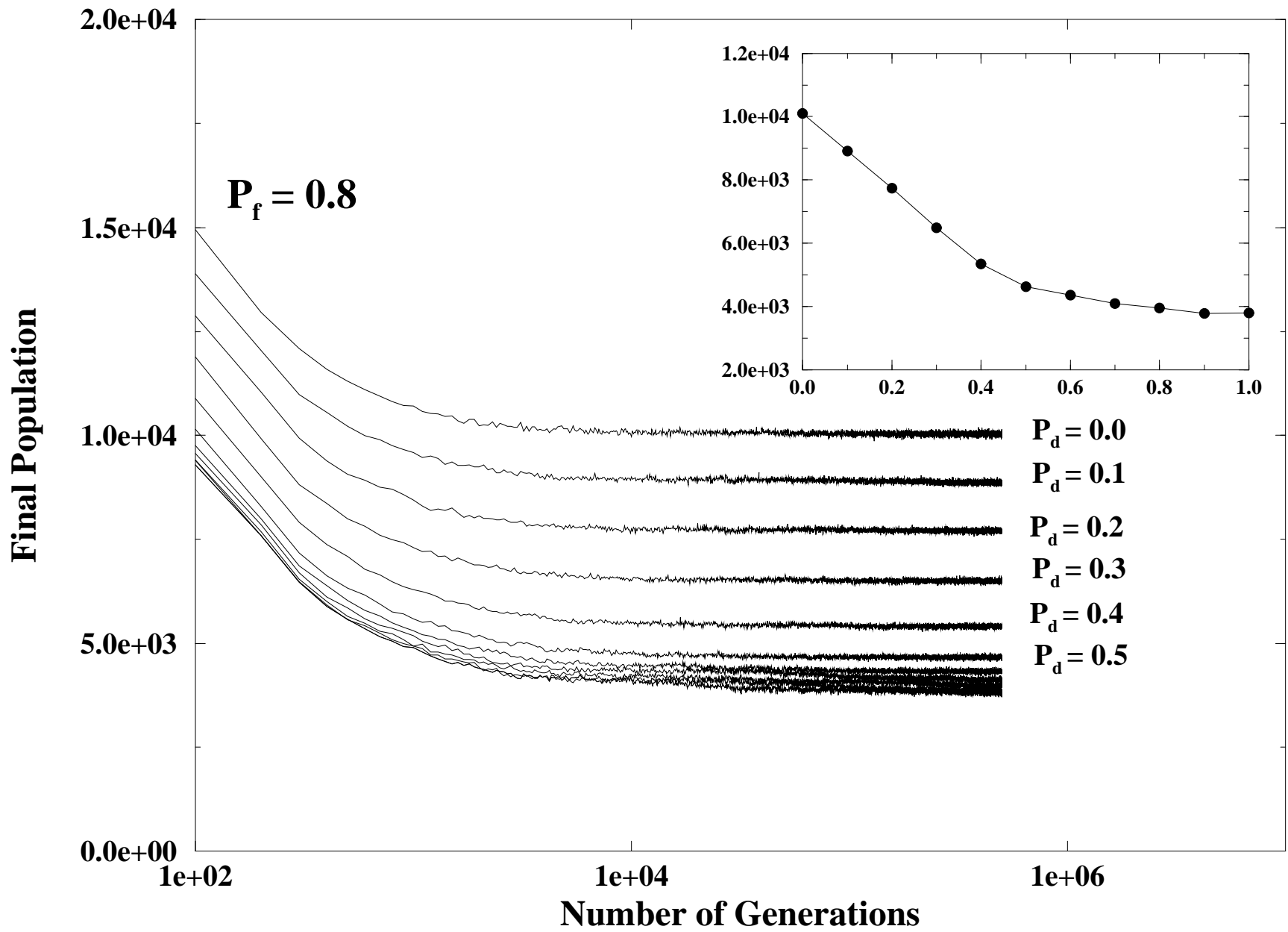
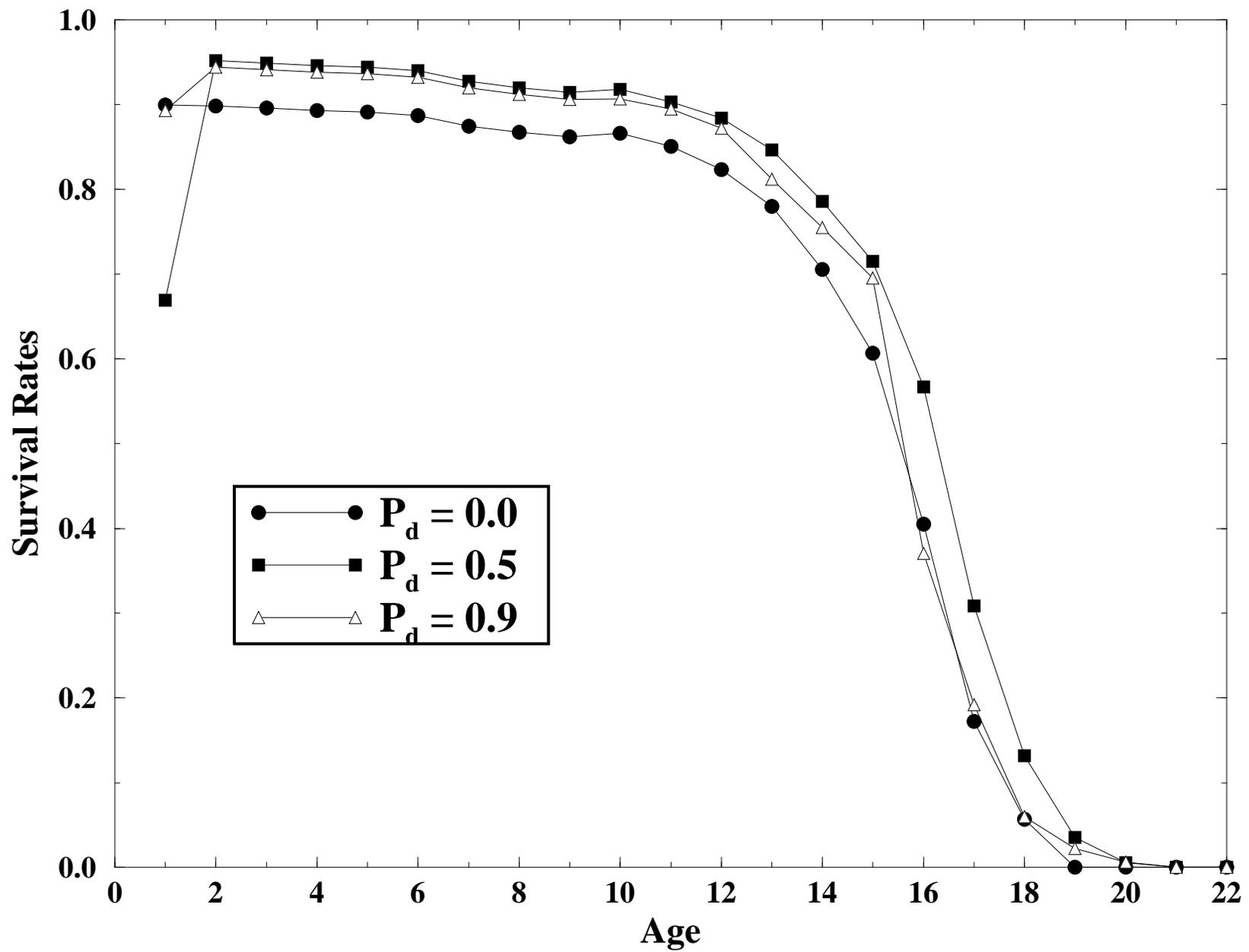
\end{document}